\documentclass{aa}
\usepackage{graphics,epsfig}
\usepackage{verbatim,multirow}
\usepackage{natbib}
\usepackage{amssymb,amsmath}
\usepackage{xspace}
\usepackage{natbib}
\newcommand{\Ha}{H$\alpha$\xspace}
\newcommand{\Hb}{H$\beta$\xspace}
\newcommand{\Hg}{H$\gamma$\xspace}
\newcommand{\LHa}{$L_{\mathrm{H}\alpha}$\xspace}
\newcommand{\logOH}{$12+\log(\mathrm{O}/\mathrm{H})$\xspace}
\newcommand{\kms}{\,km\,s$^{-1}$\xspace}
\newcommand{\HII}{\ion{H}{ii}\xspace}
\newcommand{\OIII}{[\ion{O}{iii}]}
\newcommand{\OII}{[\ion{O}{ii}]}
\newcommand{\NII}{[\ion{N}{ii}]}
\newcommand{\dV}{$\Delta V_\mathrm{max}$}
\begin{document}

\title{Tidal Dwarf Candidates in a Sample of Interacting Galaxies}
\subtitle{II.~Properties and Kinematics of the Ionized
  Gas\thanks{Based on observations collected at the European Southern
    Observatory, La Silla, Chile (ESO No 64.N-0361).}$^{,}$\thanks{Figs.~8 to 20, Table 5 and Appendices A and B are published online only.}
}
\author{P.~M.~Weilbacher\inst{1} 
  \and
  P.-A.~Duc\inst{2} 
  \and
  U.~Fritze-v.~Alvensleben\inst{1}
} 
\authorrunning{Weilbacher et al.}
\titlerunning{TDG Candidates in a Sample of Interacting Galaxies}
\offprints{P.~Weilbacher}
\institute{Universit\"ats-Sternwarte, Geismarlandstr.~11, 37083 G\"ottingen, Germany,\\
  \email{\{weilbach,ufritze\}@uni-sw.gwdg.de} 
  \and 
  CNRS URA 2052 and CEA, DSM, DAPNIA, Service d'Astrophysique,
  Centre d'Etudes de Saclay, 91191 Gif-sur-Yvette Cedex, France, \email{paduc@cea.fr} 
} 
\date{Received 3 May 2002; Accepted 17 Oct 2002}
\abstract{We present low-resolution spectroscopy of the ionized gas in
  a sample of optical knots located along the tidal features of 14
  interacting galaxies and previously selected as candidates of Tidal
  Dwarf Galaxies (TDGs).  From redshift measurements, we are able to
  confirm their physical association with the interacting system in
  almost all cases. For most knots, the oxygen abundance does not
  depend on the blue luminosity. The average, \logOH$=8.34\pm0.20$, is
  typical of TDGs and comparable to that measured in the outer stellar
  disk of spirals from which they were formed. A few knots showing low
  metallicities are probably pre-existing low-mass companions.  The
  estimated \Ha luminosity of the TDG candidates is higher than the
  one of typical individual \HII regions in spiral disks and
  comparable to the global \Ha luminosity of dwarf galaxies.
  We find several instances of velocity gradients with amplitudes
  apparently larger than 100\kms in the ionized gas in the tidal knots
  and discuss various possible origins for the large velocity
  amplitudes. While we can exclude tidal streaming motions and
  outflows, we cannot rule out projection effects with the current
  resolution. The velocity gradients could be indicative of the
  internal kinematics characteristic of self-gravitating objects.
  Higher resolution spectra are required to confirm whether the tidal
  knots in our sample have already acquired their dynamical
  independence and are therefore genuine Tidal Dwarf Galaxies.
  \keywords{Galaxies: formation -- Galaxies: interactions -- Galaxies:
    optical spectroscopy -- Galaxies: photometry} }
\maketitle
%
%
\section{Introduction}\label{sec:introEFOSC}
The formation of Tidal Dwarf Galaxies (TDGs) in interacting galaxies is
by now a well recognized phenomenon.  Previous work on the subject has
mostly focused on detailed, multi-wavelength analyses of individual
systems \citep[e.g.][etc.]{HGvG+94,DBW+97}. The first attempt to
create a large sample of TDGs was presented by \citet[][hereafter {\it
Paper~I}]{WDF+00}, how we selected TDG candidates among optical knots
in the tidal features of 10 interacting systems situated at different
redshifts below $z=0.1$.  The parent galaxies were selected from the
catalog of \citep{AM87} to resemble the perturbed systems observed in
deep surveys of the distant universe.  Evolutionary synthesis modeling
was used to rule out background objects by their broad-band colors.

What makes an optical clump in tidal features a Tidal Dwarf Galaxy is
still not well-defined.  In several studies, the most luminous knots
in or near tidal tails were classified as TDGs. However, contamination
by background galaxies is not unlikely. For instance, the blue object
of apparent dwarf galaxy luminosity positioned at the tip of one of
the long tidal tails in the Superantennae, presented by \citet{MLM91}
as a TDG candidate, was found to be a background galaxy (F.~Mirabel,
priv.~comm.).
Another example is the study of \citet{HCZ96}, who found numerous
TDG candidates in the eastern tidal tail of NGC\,7319, a member of
the Stephan's Quintet (HCG\,92). Other studies could only confirm the
association of one of these with the compact group \citep{XST99,IV01}.
galaxy-sized accumulations of material.  As normal dwarf galaxies are
stable entities with their own dynamics, the best definition of a TDG
is that it be a self-gravitating entity \citep{DBS+00,WD01}. Thus, to
confirm a knot in a tail as a genuine TDG, the velocity distribution
within the knot has to be measured, to determine if it is decoupled
from the expanding motion of the tidal tail, and possibly rotating,
obviously a difficult task.  Studying the dynamics of the HI gas,
\citet{HvdH+01} {\it failed} to prove that the most well known tidal
dwarf galaxy candidate, discovered by \citet{Sch78} and \citet{MDL92}
in the southern tail of the Antennae, is actually gravitationally
bound.

In this paper, we present the spectrophotometric follow-up of the
photometric survey of interacting galaxies presented in Paper~I. This
is the first spectroscopic investigation of a {\it sample} of TDG
candidates. The immediate objective of this work is to confirm the
physical association of the observed tidal knots with the parent
interacting system. This enables us to judge the effectiveness of a
preselection of star-forming TDG candidates based only on photometric
models such as the ones used in Paper~I.  We also obtain some key
input parameters for the evolutionary synthesis modeling code, such as
extinction, metallicity, and Balmer line equivalent widths.  We use
this data as input for the models in Paper~III of this series
(Weilbacher et al.~in prep.) to determine the evolutionary status of
the TDG candidates. Here, we also make a first attempt to probe the
dynamical status and hence real nature of the tidal objects using
long-slit velocity curves.

After describing our data acquisition and analysis methods in
Sect.~\ref{sec:obsEFOSC}, we present the basic results for our sample of
TDG candidates in Sect.~\ref{sec:resEFOSC}. In Sect.~\ref{sec:discEFOSC}, we
discuss these results in detail and describe our method of selecting
real TDGs from the sample using spectroscopic data. A summary of the
results is given in Sect.~\ref{sec:conclEFOSC}.  We present the photometric
data of four new interacting systems in the appendix
(App.~A) along with individual notes on the
spectroscopic results (App.~B).
%
%
\section{Observations and data analysis}\label{sec:obsEFOSC}
\begin{table}[tbp]
  \begin{center}
    \caption{Observing summary}
    \label{tab:ObsSumm}
    \begin{tabular}{ c  l l }
      \hline
      Object       & Observing Mode      & Exposure time                     \\
      \hline
      AM\,0529-565 & MOS                 & 4$\times$1200\,s            \\
      AM\,0537-292 & 2$\times$MOS        & 3$\times$1200\,s each       \\
      AM\,0547-244 & MOS                 & 3$\times$1200\,s            \\
      AM\,0547-474 & Imaging $B$,$V\!$,$R$ & 500,400,300\,s            \\
                   & 2$\times$Long-slit  & 3$\times$1200,3$\times$480s \\
      AM\,0607-444 & MOS                 & 3$\times$1200\,s            \\
      AM\,0748-665 & MOS                 & 3$\times$1200\,s            \\
      AM\,1054-325 & MOS                 & 2$\times$1200\,s            \\
                   & Long-slit           & 2$\times$600\,s             \\
      AM\,1159-530 & Imaging $B$,$V\!$,$R$ & 500,400,300\,s            \\
                   & MOS                 & 3$\times$1200\,s            \\
      AM\,1208-273 & MOS                 & 3$\times$1200\,s            \\
      AM\,1237-364 & Imaging $B$,$V\!$,$R$ & 3$\times$200,140,110s     \\
                   & MOS                 & 3$\times$1200\,s            \\
      AM\,1324-431 & Imaging $B$,$V\!$,$R$ & 600,480,300s              \\
                   & MOS                 & 3$\times$1200\,s            \\
      AM\,1325-292 & Long-slit           & 3$\times$600\,s             \\
      AM\,1353-272 & MOS                 & 3$\times$1200\,s            \\
      \hline
    \end{tabular}
  \end{center}
\end{table}
%
%
\subsection{Spectroscopic observations}
Our sample consists of the 10 interacting systems studied in Paper~I plus
four additional objects extracted from the Arp \& Madore \emph{Catalogue
of Southern Peculiar Galaxies} (see App.~A).

We observed the 14 interacting systems in January 2000 with EFOSC2
installed on the ESO 3.6m telescope (see Table~\ref{tab:ObsSumm}). We
made use both of the multi-object spectroscopic mode (MOS) with
1\farcs7 slitlets and of the long-slit mode with a 1\farcs5 width
slit. The geometry and orientations of all slits are indicated in
Figs.~8 to 20.
The finding charts consist of deep EFOSC2 $R$-band images (logarithmic).

The low resolution grism \#11 was used (13.2\,\AA\ FWHM resolution for a
1\arcsec\ slit), providing a spectral coverage from 3380 to 7540\,\AA\ 
for a central slit.  The detector was a LORAL CCD with 15\,$\mu$m
pixel size (0.157\arcsec on the sky), read out binned 2$\times$2.

The weather was photometric; several spectroscopic standard stars were
observed per night. The seeing varied slightly between 0\farcs85 and
1\farcs20. For our four new systems (AM\,0547-474, AM\,1159-530,
AM\,1237-364, and AM\,1324-431, see App.~A) we also
obtained broad band images in $B$,$V\!$,$R$ and observed photometric
standard stars. The reduction in these cases was done using the same
procedure as in Paper~I. The errors in the photometric calibration are
below 0.05\,mag in all cases.

\begin{figure}
  \begin{center}
    \resizebox{\hsize}{!}{
      \includegraphics{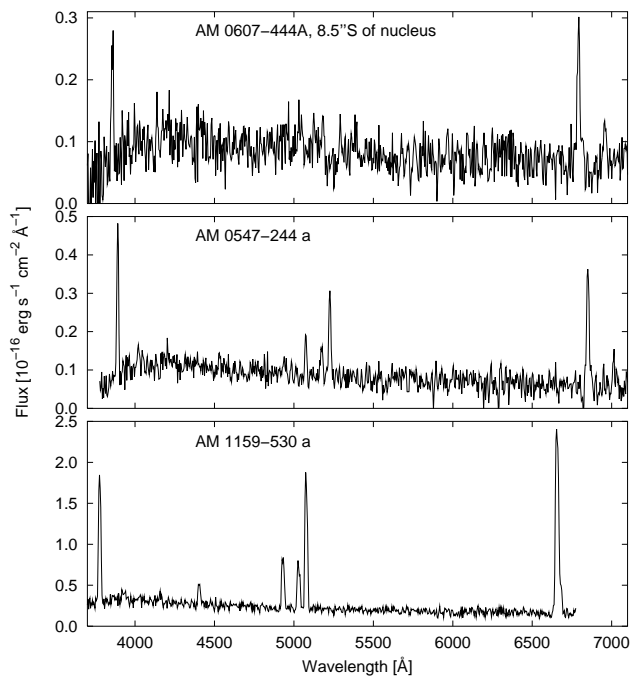}
    }
    \caption{Typical low, medium, and high S/N spectra of three of
      the observed knots. The spectra are shown with the observed 
      wavelengths and without reddening correction.  
    }
    \label{fig:sampSpec}
  \end{center}
\end{figure}

The standard reduction was done in IRAF\footnote{IRAF is written and
  supported by the IRAF programming group of the National Optical
  Astronomy Observatories (NOAO).}. We have created an IRAF task {\tt
  mosx} to handle MOS frames semi-automatically. The task proceeds in
the following manner:
\begin{enumerate}
\item extraction of the science slit spectra, flat-fields, and
  calibration HeAr spectra
\item correction of science spectra using flat-field response function
\item identification of HeAr lines and first-order wavelength calibration
\item higher order two-dimensional fit to the HeAr lines to correct
  for curved slits and the non-linear dispersion
\item flux-calibration of the individual science spectra
\end{enumerate}
A similar procedure was used to correct the long-slit spectra.
Regions of interest were then extracted from the 2D spectra and the
visible emission lines were measured using IRAF's {\tt splot} task.
Three spectra representative for those with low, medium, and high S/N
of our sample are shown in Fig.~\ref{fig:sampSpec}.

We corrected the Balmer emission line fluxes for contributions of
underlying stellar absorption. If the absorption was detected around
the emission line, we fitted two Gaussians with different widths to
deblend absorption and emission components of the lines.  If the
absorption was not visible, the we assumed a constant equivalent width
of 2\,\AA\ for the absorption, a standard value used by numerous
authors \citep[see e.g.][]{ZKH94, vZS+98}. The total reddening was
determined from the Balmer decrement, and then applied to all line
ratios to derive the corrected flux $I$ using $I/I_{\mathrm{H}\beta} =
F/F_{\mathrm{H}\beta}\,10^{\,C\,f(\lambda)}$ with the reddening curve
$f(\lambda)$ taken from \citet{Whi58}.  In absence of good S/N Balmer
emission lines, we used the Galactic extinction values of
\citet{SFD98} as given in NED\footnote{The NASA/IPAC Extragalactic
  Database (NED) is operated by the Jet Propulsion Laboratory,
  California Institute of Technology.}.
%
%
\subsection{Velocity measurements}\label{sec:veloMeasEFOSC}
We derived central velocities of all objects along the slits by
calculating weighted mean redshifts from measured emission lines,
afterwards converted into heliocentric velocities. Distances to the
parent galaxies calculated using $H_0=75$\kms\,Mpc$^{-1}$ are given in
Table~\ref{tab:PropParents} together with the absolute blue
magnitudes.

\renewcommand{\multirowsetup}{\centering}
\begin{table}[tbp]
  \begin{center}
    \caption{Distance and absolute blue magnitude of the parent galaxies}
    \label{tab:PropParents}
    \begin{tabular}{c c | c c}
      System          & D [Mpc]           & ID   &  M$_B$ [mag] \\
      \hline\hline
      \multirow{4}{19mm}{AM\,0529-565}    & \multirow{4}{8mm}{59.7} & A    &  -18.1 \\
                      &                    & (Aw) &  -15.1 \\ 
                      &                    & (Ae) &  -15.2 \\
                      &                    & D    &  -15.4 \\
      \hline
      \multirow{2}{19mm}{AM\,0537-292}    & \multirow{2}{8mm}{52.0} & A    &  -18.6 \\
                      & & B    &  -17.0 \\
      \hline
      \multirow{2}{19mm}{AM\,0547-244}    & \multirow{2}{8mm}{176}& A    &  -20.6 \\
                      & & B    &  -18.3 \\
      \hline
      \multirow{2}{19mm}{AM\,0547-474}    & \multirow{2}{8mm}{203}& A    &  -21.0 \\
                      && C    &  -19.5 \\
      \hline
      \multirow{2}{19mm}{AM\,0607-444}    & \multirow{2}{8mm}{170}& A    &  -20.8 \\
                      && (b)  &  -16.0 \\
      \hline
      \multirow{1}{19mm}{AM\,0642-325}    & \multirow{1}{8mm}{366}& A    &  -21.4 \\
      \hline
      \multirow{2}{19mm}{AM\,0748-665}    & \multirow{2}{8mm}{313}& A    &  -19.5 \\
                      && B    &  -21.1 \\
      \hline
      \multirow{2}{19mm}{AM\,1054-325}    & \multirow{2}{8mm}{52.9}& A    &  -19.1 \\
                      && B    &  -18.3 \\
      \hline
      \multirow{1}{19mm}{AM\,1159-530}    & \multirow{1}{8mm}{60.7}& A    &  -20.2 \\
      \hline
      \multirow{2}{19mm}{AM\,1208-273}    & \multirow{2}{8mm}{166}& A    &  -20.5 \\
                      && C    &  -17.1 \\
      \hline
      \multirow{2}{19mm}{AM\,1237-364}    & \multirow{2}{8mm}{76.1}& A    &  -19.9 \\
                      && B    &  -16.1 \\
      \hline
      \multirow{2}{19mm}{AM\,1324-431}    & \multirow{2}{8mm}{142}& A    &  -19.9 \\
                      && B    &  -18.1 \\
      \hline
      \multirow{2}{19mm}{AM\,1325-292}    & \multirow{2}{8mm}{60.1}& A    &  -20.0 \\
                      && B    &  -20.5 \\
      \hline
      \multirow{2}{19mm}{AM\,1353-272}    & \multirow{2}{8mm}{159}& A    &  -20.2 \\
                      && B    &  -18.1 \\
      \hline
    \end{tabular}
  \end{center}
\end{table}

Of the 36 objects selected as TDG candidates in Paper~I, we obtained
spectrophotometric data for 30 and determined the redshifts of 24.
The inferred velocities are very close to those of the parent systems
(mean velocity difference of 80\kms, and never exceeding a difference
of 350\kms) and therefore indeed imply association. For six objects
where we did not detect emission lines (despite their blue color) no
redshift could be measured. Their low surface brightnesses did not
allow us to detect any absorption lines, either.

Among the 7 new TDG candidates presented in this paper (see
App.~A),
four exhibit emission lines at the same velocity as the main system, one
has a featureless spectrum, and two were not observed spectroscopically.

Whenever gradients were visible on the 2D spectra, we tried to derive
the velocity profiles along the slit by fitting the lines with
Gaussians using the IRAF task {\tt fitprofs}. We used the HeAr line
closest to the position in the corresponding wavelength calibration
frame to correct for residual distortion.  The velocity profiles
obtained along a given slit with the brightest emission lines were
then compared. Those that were inconsistent or had too large
fitting-errors were excluded from the analysis.  The final velocity
curve was obtained by averaging the line velocities measured for each
pixel.

Note that, given the pixel scale and seeing, the velocity measurements
along the slit are not independent. Our low resolution spectra did not
allow us to get a precision better than $\pm 30$\kms for the combined
fits of lines with the highest S/N.
%
%
\subsection{Oxygen abundance measurements}\label{sec:oxygen}
\begin{figure}
  \centering
  \resizebox{\hsize}{!}{
    \includegraphics{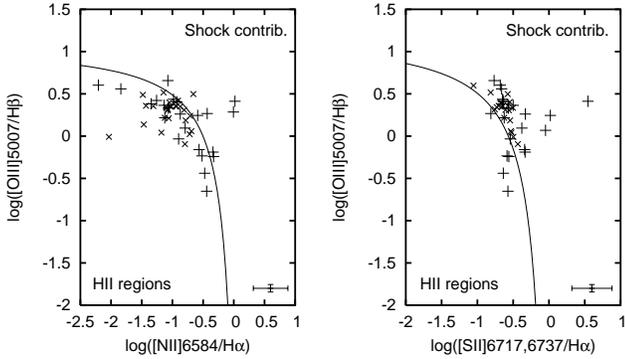}
  }
  \caption{Physical properties of the emission line regions in
    our sample. The flux line ratios are indicated. The solid lines
    give the limit for ionization by a zero age starburst
    \citep[following][]{DKH+00}. Large pluses are regions in the
    parent galaxies, small crosses denote the knots. The typical
    error for the data points is shown in the lower right corner.
  }
  \label{fig:shocks}
\end{figure}

The vast majority of the emission line regions detected in our spectra
are characterized by flux line ratios typical of \HII regions (see
Fig.~\ref{fig:shocks}). A clear substantial contribution by shock
ionization \citep{DKH+00} is only observed towards several regions of
AM\,1054-325A, including its nucleus.

We estimated the oxygen abundances in all confirmed \HII regions using
several methods proposed in the literature.  For a few objects with
a tentative detection of the faint, temperature sensitive, \OIII4363
line, we could get a physical determination of the oxygen abundance
\citep{Shi90}.  However, because of its low S/N and uncertain deblending
with the close \Hg line, the resulting value can only be regarded as
a lower limit to the actual abundance in most cases.  We therefore had
to rely on empirical methods such as the popular $R_{23}$, which uses
the measurement of the lines \OIII5007+\OIII4959+\OII3727 normalized
to the \Hb flux. The error of this method is of the order of 0.1\,dex
\citep{Pil00}. A similar method is the ``excitational'' method of
\citet{EP84}, which empirically relates the ratio \OIII5007+\OIII4959 to \Hb to
the oxygen abundance.  Unfortunately, both methods are degenerate:
for a given value of $R_{23}$, two abundances are possible. In order to
choose between the lower and higher values, we relied on the \NII/\OII\
indicator \citep{vZS+98}. \citet{Pil00,Pil01} recently proposed a new
method based on a complex combination of the \OIII\ and \OII\ line
fluxes called the ``$P$-method''. This calibration better matches the
the oxygen abundances determined with the physical method, especially
for high metallicity \HII regions.
We assigned a systematic error of 0.05\,dex to the $P$-method for the
upper branch of the $P$-[O/H], and a slightly larger error of
0.08\,dex for the lower branch (called $P_3$).  Finally the
$N$-calibration proposed by \citet{vZS+98}, based on the flux-ratio of
\Ha to \NII6548,6584, provides another empirical method of deriving
the oxygen abundance, but suffers from deblending problems in our low
resolution spectra; its uncertainty is as high as 0.2\,dex.

We compared the values obtained by each of these methods and checked
the consistency of the results. Because of its smaller intrinsic
uncertainty, we decided to preferentially opt for the value given by
the $P$-method unless the observed parameter $P$ was outside its
validity range.  The $N$-calibration was only used when no other
methods yielded reliable results. The final oxygen abundances are
listed in Table~\ref{tab:lineResults}. The errors take into account
the errors of the flux measurements plus the systematic errors of the
calibrations.

\begin{table*}[tbp]
  \begin{center}
    \caption{Line ratios, extinction values, and oxygen abundances of all observed knots.}
    \label{tab:lineResults}
    \begin{tabular}{l | c c | c c c c | l }
       knot & $F$(\Hb)  & C$^1$ & \OII3727 & \OIII5007 & \NII6584 & [\ion{S}{ii}]6717,6731 & Oxygen Ab.$^2$ \\
            & [$10^{-15}$\,erg\,s$^{-1}$\,cm$^{-2}$] & & \multicolumn{4}{|c|}{[$I$(line)/$I$(\Hb)]} & \logOH \\
       \hline
 AM\,0529-565D                   & 3.61$\pm$0.05 & 0.07 & --            & 3.97$\pm$0.28 & --            & 0.24$\pm$0.02 & {\bf 7.61}$\pm$0.20 (ex)      \\
 AM\,0529-565e                   & 0.12$\pm$0.04 & 0.59 & 6.19$\pm$0.38 & 2.29$\pm$0.62 & 0.14$\pm$0.28 & --            &       8.39$\pm$0.16 (ex)      \\
 AM\,0537-292a                   & 0.86$\pm$0.03 & 0.05 & 3.25$\pm$0.05 & 2.30$\pm$0.05 & 0.38$\pm$0.03 & 0.89$\pm$0.04 &       8.32$\pm$0.05 (P)       \\
 AM\,0537-292\,(5\farcs0\,S\,of\,b)&0.11$\pm$0.03& 0.05 & 2.29$\pm$0.20 & 1.73$\pm$0.20 & 0.41$\pm$0.25 & --            &       8.50$\pm$0.07 (P)       \\
 AM\,0537-292c                   & 1.25$\pm$0.04 & 0.68 & 3.63$\pm$0.04 & 2.29$\pm$0.04 & 0.34$\pm$0.04 & 0.79$\pm$0.09 &       8.26$\pm$0.05 (P)       \\
 AM\,0537-292c (N)               & 0.77$\pm$0.03 & 0.50 & 3.42$\pm$0.04 & 2.03$\pm$0.05 & 0.26$\pm$0.06 & 0.87$\pm$0.12 &       8.31$\pm$0.05 (P)       \\
 AM\,0537-292d                   & 0.35$\pm$0.03 & 0.45 & 4.27$\pm$0.14 & 2.17$\pm$0.20 & 0.31$\pm$0.11 & 1.18$\pm$0.35 &       8.20$\pm$0.06 (P)       \\
 AM\,0537-292g                   & 0.53$\pm$0.04 & 0.44 & 3.80$\pm$0.09 & 2.27$\pm$0.10 & 0.14$\pm$0.10 & 0.63$\pm$0.18 &       8.24$\pm$0.05 (P)       \\
 AM\,0537-292h                   & 0.11$\pm$0.05 & 0.48 & 5.44$\pm$0.41 & --            & 0.28$\pm$0.28 & --            &       8.23$\pm$0.39 (N)       \\
 AM\,0547-244A,bright\,knot      & 1.92$\pm$0.06 & 0.19 & 2.27$\pm$0.03 & 1.14$\pm$0.04 & 0.54$\pm$0.03 & 0.77$\pm$0.04 &       8.52$\pm$0.05 (P)       \\
 AM\,0547-244A,E\,tail           & 0.17$\pm$0.03 & 0.05 & 2.41$\pm$0.11 & 0.98$\pm$0.12 & 0.02$\pm$0.03 & 0.78$\pm$0.21 &       8.50$\pm$0.06 (P)       \\
 AM\,0547-244A,outer\,E\,tail    & 0.14$\pm$0.03 & 0.10 & 1.77$\pm$0.16 & 1.37$\pm$0.16 & 0.09$\pm$0.06 & --            &       8.60$\pm$0.07 (P)       \\
 AM\,0547-244a                   & 0.17$\pm$0.03 & 0.68 & 3.94$\pm$0.12 & 2.04$\pm$0.21 & 0.45$\pm$0.04 & 0.72$\pm$0.23 &       8.23$\pm$0.06 (P)       \\
 AM\,0547-244b                   & 0.39$\pm$0.03 & 0.83 & 4.86$\pm$0.09 & 2.53$\pm$0.10 & 0.35$\pm$0.04 & 0.89$\pm$0.30 &       8.39$\pm$0.15 (ex)      \\
 AM\,1054-325g                   & 1.72$\pm$0.08 & 0.25 & 2.05$\pm$0.03 & 2.22$\pm$0.03 & 0.28$\pm$0.02 & 0.64$\pm$0.05 &       8.48$\pm$0.05 (P)       \\
 AM\,1054-325h (N)               & 0.64$\pm$0.05 & 0.41 & 3.39$\pm$0.09 & 2.07$\pm$0.16 & 0.25$\pm$0.05 & 0.52$\pm$0.07 &       8.28$\pm$0.05 (P)       \\
 AM\,1054-325h (S)               & 0.32$\pm$0.05 & 0.29 & 2.92$\pm$0.14 & 2.24$\pm$0.17 & 0.37$\pm$0.13 & 0.57$\pm$0.09 &       8.24$\pm$0.06 (P)       \\
 AM\,1054-325i                   & 1.79$\pm$0.07 & 0.44 & 2.22$\pm$0.03 & 3.28$\pm$0.05 & 0.21$\pm$0.04 & 0.45$\pm$0.06 & {\bf 7.76}$\pm$0.10 (R$_{23}$)\\
 AM\,1054-325j                   & 4.37$\pm$0.07 & 0.32 & 3.08$\pm$0.02 & 2.28$\pm$0.02 & 0.29$\pm$0.02 & 0.63$\pm$0.03 &       8.55$\pm$0.05 (P)       \\
 AM\,1054-325o                   & 7.82$\pm$0.12 & 0.47 & 2.41$\pm$0.01 & 2.20$\pm$0.01 & 0.24$\pm$0.01 & 0.58$\pm$0.01 &       8.45$\pm$0.05 (P)       \\
 AM\,1159-530a (E)               & 0.87$\pm$0.03 & 0.78 & 2.71$\pm$0.05 & 2.23$\pm$0.06 & 0.46$\pm$0.05 & --            &       8.40$\pm$0.05 (P)       \\
 AM\,1159-530a (W)               & 1.28$\pm$0.05 & 0.56 & 2.58$\pm$0.04 & 2.63$\pm$0.07 & 0.43$\pm$0.04 & --            &       8.40$\pm$0.05 (P)       \\
 AM\,1159-530q                   & 0.39$\pm$0.05 & 1.03 & 2.27$\pm$0.11 & 1.54$\pm$0.17 & 0.81$\pm$0.11 & 1.41$\pm$0.31 &       8.51$\pm$0.06 (P)       \\
 AM\,1208-273e                   & 0.22$\pm$0.03 & 1.75 & 3.54$\pm$0.15 & 0.80$\pm$0.15 & 0.42$\pm$0.06 & 0.96$\pm$0.30 &       8.33$\pm$0.07 (P)       \\
 AM\,1208-273h                   & 0.28$\pm$0.04 & 0.87 & 3.64$\pm$0.14 & 1.32$\pm$0.14 & 0.26$\pm$0.03 & 0.67$\pm$0.17 &       8.31$\pm$0.06 (P)       \\
 AM\,1208-273i                   & 0.21$\pm$0.02 & 0.13 & 2.51$\pm$0.11 & 1.10$\pm$0.09 & 0.17$\pm$0.03 & 0.77$\pm$0.15 &       8.48$\pm$0.06 (P)       \\
 AM\,1237-364d                   & 0.15$\pm$0.04 & 0.14 & 3.08$\pm$0.24 & 3.15$\pm$0.35 & 0.74$\pm$0.09 & 0.89$\pm$0.13 &       8.30$\pm$0.07 (P)       \\
 AM\,1237-364e                   & 0.16$\pm$0.02 & 0.14 & 2.85$\pm$0.15 & 1.65$\pm$0.20 & 0.19$\pm$0.11 & 0.60$\pm$0.15 &       8.41$\pm$0.06 (P)       \\
 AM\,1237-364g                   & 2.28$\pm$0.06 & 0.36 & 1.92$\pm$0.03 & 3.08$\pm$0.04 & 0.10$\pm$0.03 & --            & {\bf 7.72}$\pm$0.05 (phys)    \\
 AM\,1324-431c                   & 0.12$\pm$0.06 & 0.21 & 2.82$\pm$0.34 & 1.09$\pm$0.29 & 0.66$\pm$0.16 & --            &       8.66$\pm$0.22 (N)      \\
    \end{tabular}
  \end{center}
$^1$Total extinction from the Balmer decrement.\\
$^2$In brackets we list the method used to derive the oxygen abundance as described in Sect.~\ref{sec:oxygen}.
\end{table*}
%
%
\section{Results}\label{sec:resEFOSC}
The individual results for each system are summarized graphically:
Figs.~8 to 20
show the optical images together with several observed position-velocity
diagrams and derived velocity curves.  All objects and knots discussed
are labeled; the slits are overplotted in their real scale (length
and width). Measured mean velocities or redshifts are indicated.
Low metallicity objects with 12+log(O/H)$<$8.0 are circled and objects
in the slits without firmly detected lines are marked ``n.d.''. More
detailed descriptions of each system may be found in Paper~I and in
App.~B.

Below, we present the properties of the objects that proved to be
physically associated with the tidal features.  We concentrate our
analysis on the emission-line regions that they host.  Our selection
of TDG candidates was biased towards star-forming objects (see
Paper~I) and, indeed, the vast majority of them host \HII regions.
%
%
\subsection{Metallicity}\label{sec:resEFOSC:met}
\begin{figure*}[htbp]
  \begin{center}
    \resizebox{0.8\hsize}{!}{
      \includegraphics{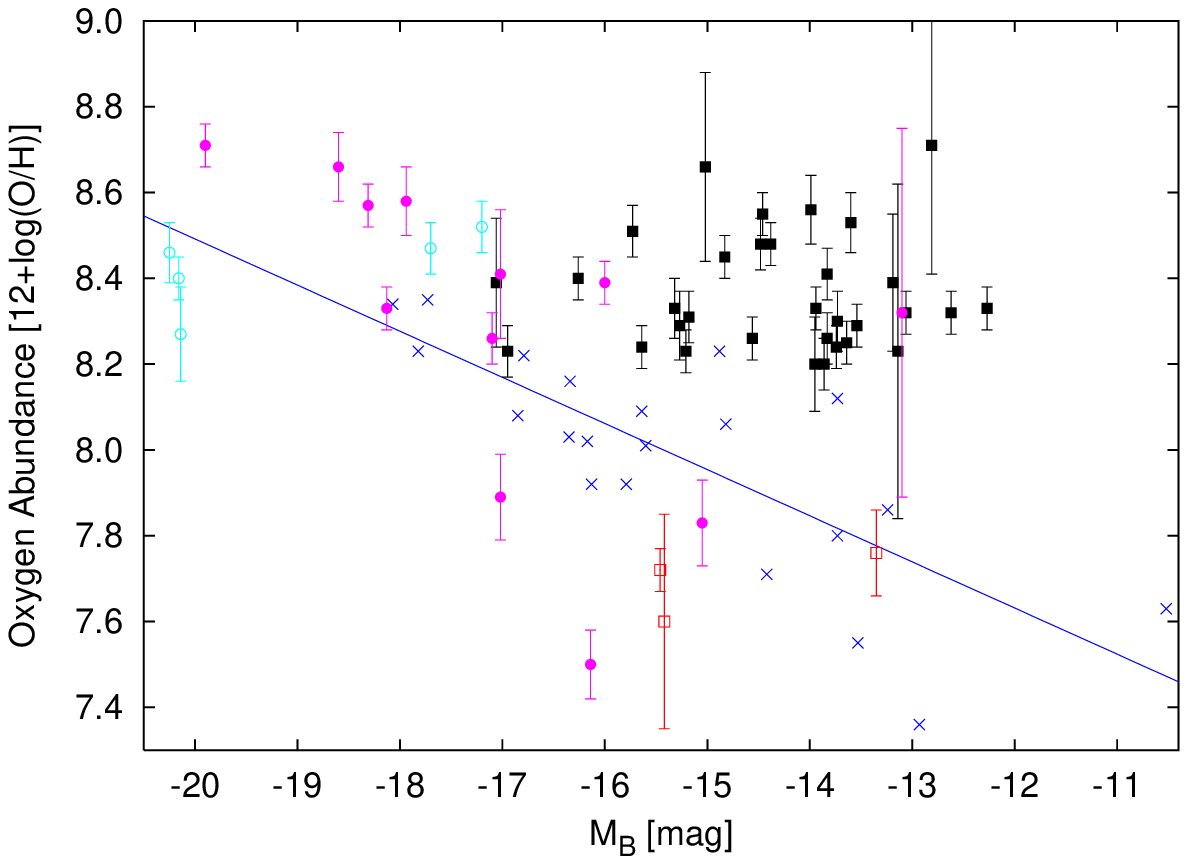}
    }
    \caption{Metallicity-luminosity relation for different types 
      of objects: 
      local isolated dwarf galaxies (crosses), 
      knots in the tidal features (filled/open squares),
      main group members (filled circles), 
      background objects (open circles).
    }
    \label{fig:met-lum}
  \end{center}
\end{figure*}

The metallicity-luminosity relation for the objects in our sample and for isolated
dwarf galaxies is shown  in Fig.~\ref{fig:met-lum}.
 Nearby
dwarf irregular galaxies \citep{RM95} follow the well-known correlation
indicated by the linear fit.  The massive galaxies in our sample --
main galaxies and background objects -- also follow this
relation\footnote{The exceptions are two low metallicity star-bursting
  dwarf galaxies (AM\,1237-364\,B and C), the western nucleus `Aw' of
  AM\,0529-565, and one low surface brightness galaxy near the
  interacting system AM\,1353-272.}.  On the other hand, knots along
the tidal features deviate significantly from the
 relation and seem to have a metallicity which is independent of 
luminosity. The bulk of the emission-line knots
in our sample have \logOH in the range $8.3\dots8.7$ (mean
\logOH$=8.34\pm0.14 \approx 1/4\,\sun$) and luminosities $M_B = -12
\dots -17$\,mag.  They occupy the position of previously published
TDGs \citep[see e.g.][]{DBS+00}.  Three objects, AM\,0529-565\,D,
AM\,1237-364\,g, and AM\,1054-325\,i, however, have oxygen abundances
at least 0.4\,dex lower than the rest of the knots.
%
%
\subsection{Internal extinction}
In knots where we could determine internal extinction from the Balmer
decrement in addition to the Galactic absorption, we derive $A_V$ in
the range $0.0\dots3.3$\,mag with a mean of $A_V = 0.81$\,mag.
 
This is roughly half the mean extinction and range observed in \HII
regions in spiral galaxies \citep[e.g.~mean $A_V = 3.1$\,mag and $0 <
A_V < 6$\,mag for \HII regions in the central part of M51,][]{SPE+01}.
This is understandable as our slits sample a much larger field
(typically several hundreds of parsecs, see Sect.~\ref{sec:discEFOSC:Halum})
than the spatially resolved HST imaging studies of Scoville et
al.~(typically a few parsecs). Our slits therefore average over
several individual \HII regions and the regions between them. As the
extinction is typically much higher within \HII regions where the dust
accompanies the star formation process, this naturally explains the
smaller range and smaller mean value we see in our sample.

Our mean extinction is also comparable to the typical $A_V \simeq
1.0$\,mag of the young star clusters in the less obscured regions of
the Antennae \citep{WZL99}.
%
%
\subsection{\Ha luminosities}\label{sec:resEFOSC:Halum}
\begin{figure}
  \centering
    \resizebox{\hsize}{!}{
      \includegraphics{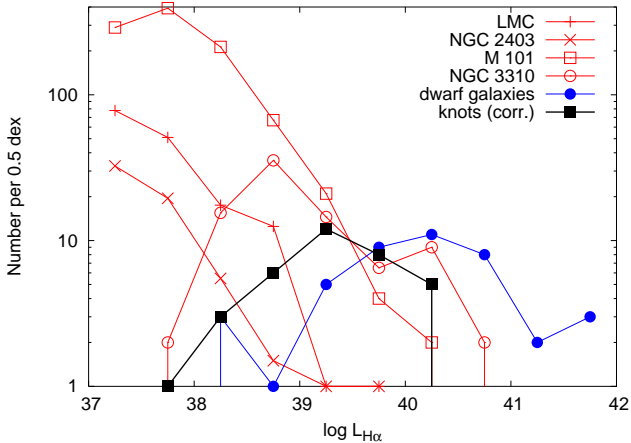}
    }
    \caption{\Ha luminosity distribution 
      of the observed knots with confirmed higher metallicity
      compared with literature values for individual
      \HII regions in spiral disks and total blue dwarf galaxies. 
    }
    \label{fig:Ha_hist_comp}
\end{figure}

Other authors before have used the \Ha luminosity to discriminate
between candidates for TDGs and normal \HII regions \citep[see
e.g.][]{IV01}.  We have  compiled the \Ha luminosities of all
the tidal knots studied in this survey.  Note that the raw
luminosities (as e.g.~presented in Table~\ref{tab:lineResults}) are
lower limits since they are measured with an aperture-limited slit and
not from narrow band imaging.  From the size of the optical knots as
measured on $B$-band images and the area covered by the slit, we have
derived correction factors between 1.5 and 5 depending on the coverage
of the knot by the slit and used these to estimate the total \Ha
luminosities of the knots. The average \Ha luminosity we derive is
$2.2\cdot10^{39}$\,erg\,s$^{-1}$.  How does this value compare with
that measured in typical \HII regions of disk galaxies and other dwarf
galaxies?

A histogram of the \Ha luminosity {\it distribution}  of the tidal knots is
shown in Fig.~\ref{fig:Ha_hist_comp} where it is compared to the
luminosity {\it function} (LF) of \HII regions in four nearby galaxies
of various morphological types (the LMC, NGC\,2403, M\,101, and
NGC\,3310). The data was extracted from the catalog of \citet{KEH89},
compiled from ground-based narrow band images. This plot shows an
apparent large diversity in the shapes of the \HII LFs.  All
individual \HII regions in the LMC and the Sc spiral NGC\,2403 have
\Ha luminosities about one order of magnitude below the most luminous
knots in our survey ($\sim 10^{40}$\,erg\,s$^{-1}$).  The spiral
M\,101 hosts a few giant \HII regions reaching this value.  The LF of
the more distant galaxy NGC\,3310 (D$\simeq$19\,Mpc) is affected by
incompleteness for $\log$\LHa$<38.5$ and has a high number of luminous
\HII regions with a distribution very similar to that of our knots.
This galaxy is very perturbed and probably has recently undergone a
merger.

In the same Figure we also present a comparison with the
{\it global} \Ha luminosity distribution of ``normal''
dwarf galaxies consisting of 11 Im and 7 dI galaxies of \citet{GHB89},
12 nearby ``amorphous'' dwarf galaxies of \citet{MMH+97}, and 12 blue
compact dwarfs of \citet{CCV+01}.  
The knots have a distribution similar to the low luminosity end of the
normal dwarfs, while there are several dwarf galaxies with
luminosities one order of magnitude higher than any of the knots in
the tidal tails.
%
%
\subsection{Kinematics}\label{sec:resEFOSC:grad}
\begin{table}
  \begin{center}
    \caption{Observed velocity gradients}
    \label{tab:veloGradSumm}
    \begin{tabular}{l | r c c c}
      TDG candidate   & \centering \dV  &  $R$    & Angle   & \multicolumn{1}{l}{Extent} \\
                      &\multicolumn{1}{c}{[km\,s$^{-1}$]}
                                           &  [kpc]  & [\degr] &    \multicolumn{1}{r}{(seeing)} \\
      \hline                                                   
      AM\,0537-292a   &  70$\dots$360$\pm$50 & ~8.10   & 43      &   1\farcs5/2\farcs7 (1\farcs04) \\
      AM\,0537-292g   &           160$\pm$50 & 12.60   & 83      &   2\farcs9 (1\farcs26)          \\
      AM\,0547-244a   &            77$\pm$61 & 23.60   &  5      &   3\farcs5  (1\farcs00)         \\
      AM\,0547-244b   & 220$\dots$485$\pm$80 & 30.90   & 85      &   2\farcs4/4\farcs4 (1\farcs00) \\
      AM\,1054-325h   & 130$\dots$420$\pm$45 & ~6.40   &  2      &   2\farcs0/3\farcs5 (1\farcs00) \\
      AM\,1159-530a   &           440$\pm$45 & 29.30   & 29      &   5\farcs5 (1\farcs04)          \\
      AM\,1353-272a$^\ast$&       343$\pm$17 & 39.30   &  0      &   3\farcs0 (1\farcs02)    \\
      AM\,1353-272b$^\ast$&        87$\pm$18 & 33.70   &  0      &   2\farcs6 (1\farcs02)    \\
      AM\,1353-272c$^\ast$&        34$\pm$14 & 31.60   &  0      &   3\farcs2 (1\farcs02)    \\
      AM\,1353-272d$^\ast$&        93$\pm$~9 & 28.10   &  0      &   4\farcs6 (1\farcs02)    \\
      AM\,1353-272k$^\ast$&        50$\pm$~7 & 22.30   &  0      &   2\farcs2 (1\farcs02)    \\
      AM\,1353-272l$^\ast$&        24$\pm$~6 & 24.70   &  0      &   2\farcs6 (1\farcs02)    \\
      AM\,1353-272m$^\ast$&        43$\pm$~8 & 28.30   &  0      &   3\farcs4 (1\farcs02)    \\ 
      \hline
      \end{tabular}
  \end{center}
  $^\ast$observed with FORS2 at the VLT \citep{WFDF02}
\end{table}
\begin{figure}[!ht]
  \centering
  \resizebox{\hsize}{!}{
    \includegraphics{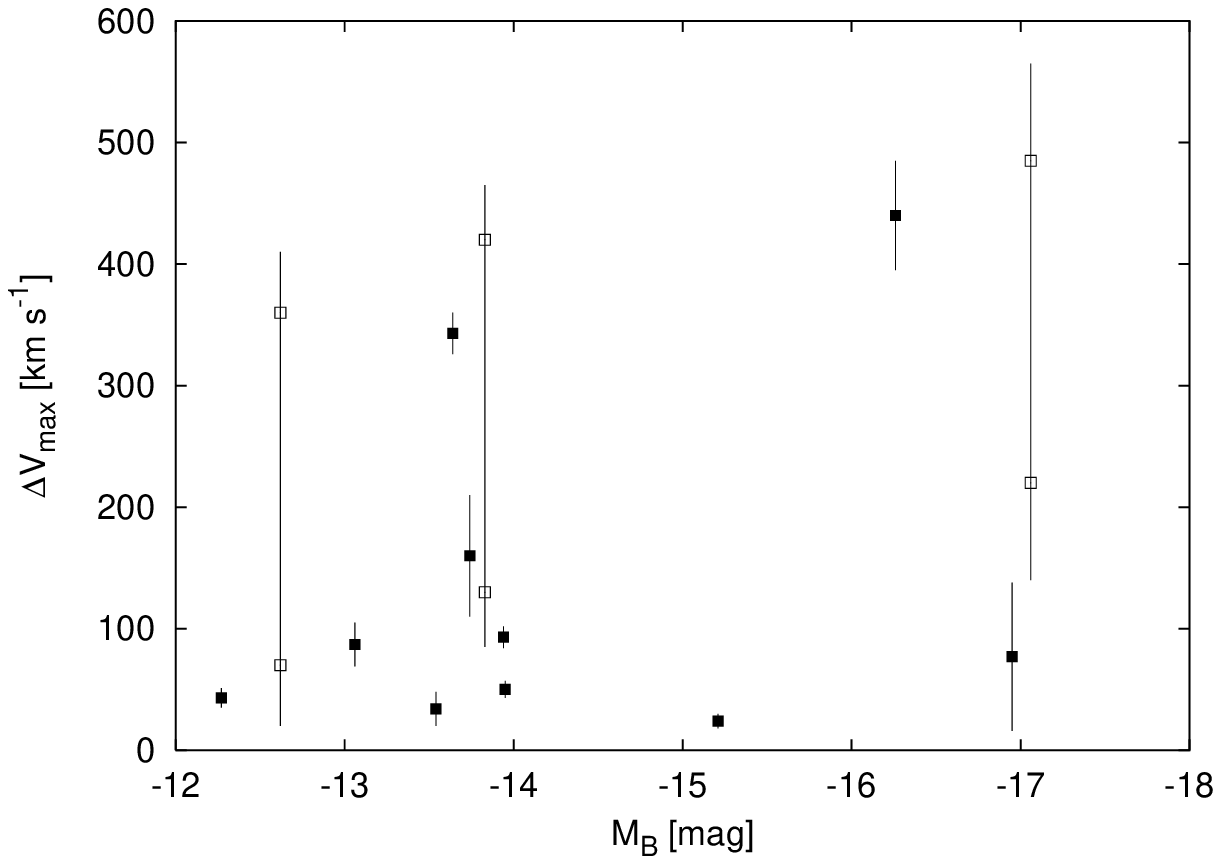}
    }
  \caption{Maximum velocity difference \dV\ vs.~absolute blue magnitude of our TDGs. 
    Filled squares mark TDGs with only one possible \dV, the three cases with
    two possible values of \dV\ have been plotted as 2 connected open squares
    to visualize the possible range in \dV.
  }
  \label{fig:dV_MB}
  \centering
  \resizebox{\hsize}{!}{
    \includegraphics{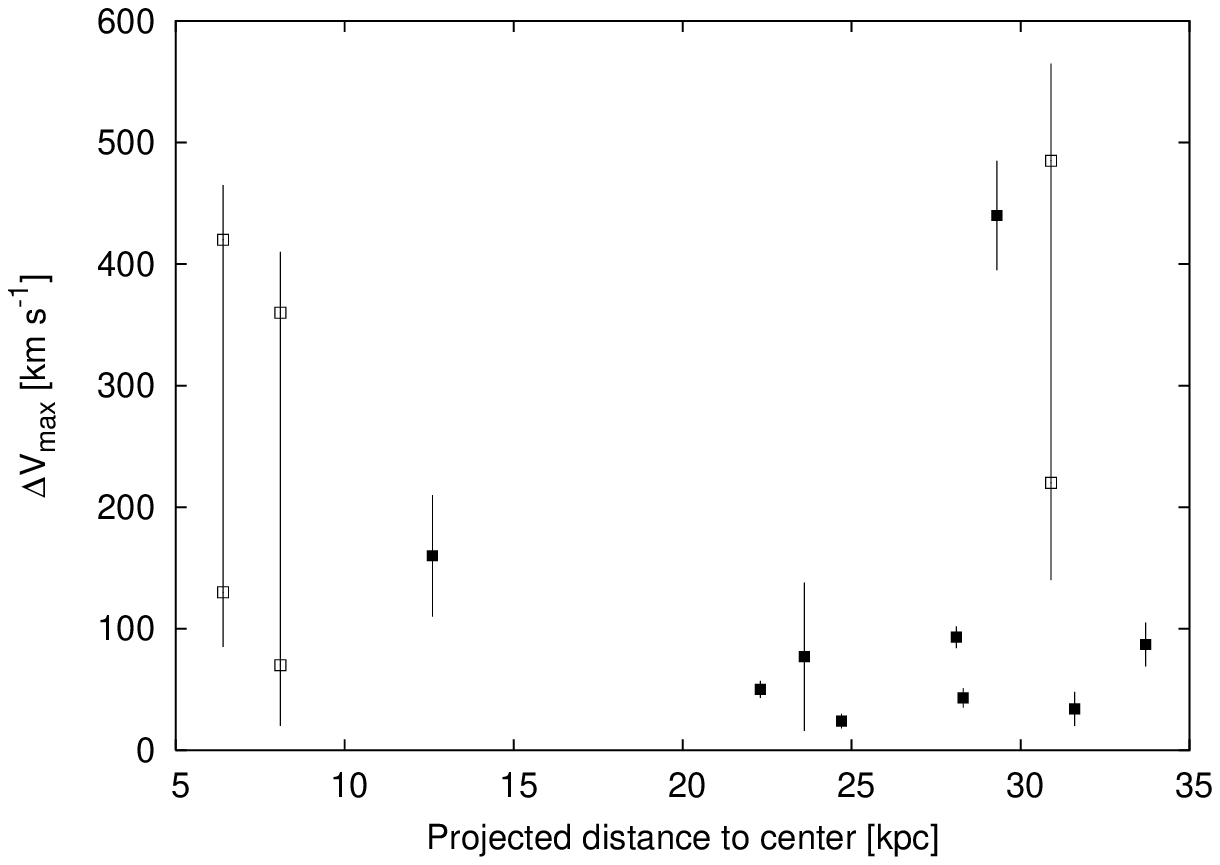}
    }
  \caption{Maximum velocity difference \dV\ vs.~projected distance 
    nucleus-knot along the tail. 
    Symbols as in Fig.~\ref{fig:dV_MB}. }
  \label{fig:dV_R}
  \centering
  \resizebox{\hsize}{!}{
    \includegraphics{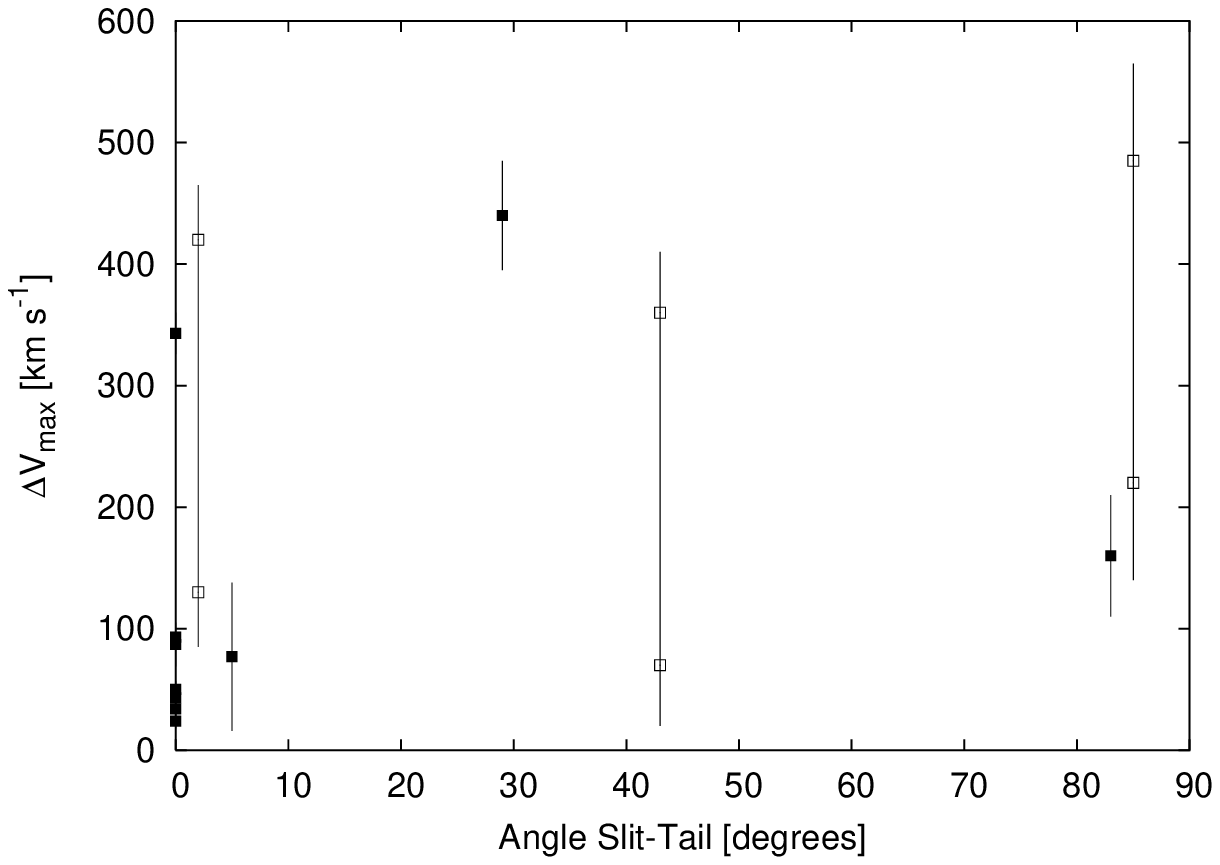}
    }
  \caption{Maximum velocity difference \dV\ vs.~angle slit-tail. 
    Symbols as in Fig.~\ref{fig:dV_MB}. }
  \label{fig:dV_Angle}
\end{figure}

Due to observational constraints, we could only partially sample the
velocity curves along the tails from our long-slit and MOS spectra. A
striking result from this study is the presence of kinematical
structures with large apparent velocity gradients along the tidal
tails.

In our low resolution data, we find several instances of knots with
discrepant velocities. In a few cases, we have enough spatial resolution
to measure apparent velocity gradients at the location of optically
detected objects: in AM\,0537-292 (Fig.~10), AM\,0547-244 (Fig.~11),
AM\,1054-325 (Fig.~15), and AM\,1159-530 (Fig.~16).
The apparent velocity differences
\dV\ between two close regions can be as high as several hundred \kms.
In three knots (AM\,0537-292a, AM\,0547-244b, and AM\,1054-325h) we
observe substructure even with our low spatial resolution. In these
cases, we note the total amplitude of the gradient and \dV\ within the
brighter knot.\\
All observed velocity gradients are summarized in
Table~\ref{tab:veloGradSumm}. There, we also list the distance $R$
between the knot in the tidal tail and the nucleus of the parent
galaxy along the tail. We also give the angle between the orientation
of the slit and the central ridge of the tail, and the extent of the
observed velocity gradient.

In Figs.~\ref{fig:dV_MB} to \ref{fig:dV_Angle}, parameters which might
affect the strength of the velocity gradient given by the \dV\ 
variable are plotted. Fig.~\ref{fig:dV_MB} shows the maximum velocity
difference versus the absolute blue magnitude.  We do not observe any
obvious trend between the stellar mass very -- roughly represented by
the blue luminosity -- and the dynamical mass to which \dV\ is
presumably a rough indicator.  In Fig.~\ref{fig:dV_R}, we see that
\dV\ of the knots does not depend on the distance to the nucleus of
the parent galaxy: small gradients occur both near to and far away
from the parent nucleus.  In Fig.~\ref{fig:dV_Angle}, we investigate
if the angle between the slit and the tail has a significant
contribution to the value of the velocity gradient. As the
observations of AM\,1353-272 with its 7 knots were performed with
slits along the tails \citep{WFDF02}, we have a strong selection
effect towards low angles, and cannot derive a dependence between the
angle and \dV. However, we note that velocity gradients are observed
at all angles between the slit and the tail.  That means that we see
velocity components both along and perpendicularly to the tail.
%
%
\section{Discussion}\label{sec:discEFOSC}
The tidal knots we observed appear to have   rather
unusual characteristics: in particular their large \Ha luminosities, and
the kinematics of their ionized gas, with velocity amplitudes greater than 100\kms
are striking. With such properties, the TDG candidates in our
sample appear to be closer to individual dwarf galaxies than to the giant HII regions
of spiral disks. 
%
%
\subsection{TDG candidates: between giant HII regions and individual dwarfs galaxies?}

\subsubsection{\Ha luminosities}\label{sec:discEFOSC:Halum}
What can we learn from the comparison with the \HII region
luminosities?  The spatial coverage of our slits is typically
3\farcs5$\times$1\farcs7 (the extraction width times the slit width
for MOS).  For the nearby systems ($D \sim 50$\,Mpc) we therefore
summed the \Ha flux over a region of 850$\times$410\,pc$^2$; for a
distance of 150\,Mpc this would increase to 2.5$\times$1.2\,kpc$^2$.
From this rough estimate is seems clear that we do not observe
individual \HII regions, but cover a region much larger in size.  With
the value found for M\,101 by \citet{PHF00} a minimum distance between
\HII regions is 60\,pc; then about 100 individual \HII regions would
``fit'' into the region we sample for our nearby systems and 1000 for
the distant ones.
With a typical individual luminosity of $10^{37}$\,erg\,s$^{-1}$
\citep{SPE+01}, one would then expect integrated luminosities of
$10^{39} \dots 10^{40}$\,erg\,s$^{-1}$, similar to the upper end of
the luminosity distribution of the knots. The luminous end of our
knots will then be the real starbursts in our sample, while the faint
end are regions forming stars at a lower level.

The comparison with the sample of dwarf galaxies is more difficult,
because several of the objects shown in Fig.~\ref{fig:Ha_hist_comp}
are not classical dwarf galaxies with $M_B < -17$\,mag but have higher
blue luminosities. The \Ha luminosity does {\it not} linearly
depend on the star formation rate for dwarf galaxies, but is strongly
affected by evolutionary effects as shown by \citet{WF01}. This does
not allow us to draw any conclusions about the star formation rates of
both types of objects. 
In summary, the \Ha luminosities emitted by the tidal knots appear to
be lower than the global \LHa of normal galaxies with $M_B \le
-19$\,mag. However, they are at least similar to and in most cases exceed
those measured in \HII regions located in the disks of nearby spiral
galaxies. It seems clear that the knots we observe are actually a
conglomerate of several individual \HII regions. If that is the case,
one may wonder whether they do belong to a single dynamical object.

\subsubsection{Metallicities}
Currently known TDGs have metallicities around 1/3 solar
\citep{DBS+00}, significantly higher than normal dwarf galaxies of
similar luminosity. This can be understood if TDGs are made out of
material which was already metal-enriched prior to the current
starburst in the progenitor galaxy: they were dynamically built from
the outer disks of their parent galaxies where the metallicity is
close to this value \citep{ZKH94}. Gas with very low metal abundance
exists in the outermost regions of spirals \citep{FGW98}.
However, the analysis carried out in Paper~I shows that our TDG
candidates are likely to contain a significant amount of old stars.
Those had to be pulled out from regions in the parent stellar disk
which were rather dense and could not be situated far beyond the
optical radius.  At their location, the metallicity should have been
close to that measured in the tidal knots, i.e.~1/4 to 1/3 solar.

Some dynamical simulations \citep[e.g.][]{BH98} show that the material
of the inner and outer parts of a tidal tail came from smaller and
larger radii of the progenitor disk, respectively. Due to the observed
metallicity gradients of spiral disks, one might expect to see
abundance gradients along tidal tails.  In the systems AM\,0537-292
and AM\,1353-272, where we could observe several knots along the
tails, however, we do not observe any systematic radial trend in their
abundances. All the knots have nearly the same oxygen abundance around
\logOH$\approx$8.34. This hints at a process which mixes material from
different regions in the progenitor galaxy and has to be checked with
new dynamical models.

The three objects with low metal abundances mentioned in
Sect.~\ref{sec:resEFOSC:met} also seem to contain a strong old stellar
component \citep{WDF+00}.  They are therefore most likely pre-existent
companion galaxies, which are either projected onto the tidal features
or falling into the interacting system.
%
%
\subsection{Velocity gradients}
What do the apparent velocity gradients tell us about the kinematics
of the knots?  Clearly such amplitudes are not normally observed in
the \HII regions of spirals where, after correction for the overall
velocity field, only 15\kms of residual vertical dispersion is
observed \citep{JB00}.  What is then the origin of such ``gradients''
in our objects? In fact, several different causes are possible.

\subsubsection{Projection effects}
Most of the velocity gradients that we observe in the knots have an
extent of 2 to 4\arcsec, i.e.~only a few times the FWHM of the typical
seeing (see column 2 of Table~\ref{tab:veloGradSumm}). The lack of
spatial resolution and related ``smoothing'' will cause physically
disconnected regions with distinct velocities to appear as being part
of a single kinematical entity.  Were that the case, the observed
velocity gradient would only be apparent and would hide projection
effects.  Some knots could in fact be external material seen in
projection onto the tail. One may also have the cases where several
knots of a bent tail or from even different tails project onto each
other.

In Sect.~\ref{sec:resEFOSC:grad} we noted three cases with two emission
line regions within one apparent knot. Although we cannot confirm
these instances as projection effects, these are candidates for this
type of velocity gradient caused by projection.

\subsubsection{Streaming motions along the tails}
The overall velocity field of tidal tails is governed by streaming
motions resulting from the gravitational tidal forces exerted during
the collision.  Identifying in them kinematically decoupled entities
is difficult when, as in our sample, the global velocity field is only
measured at discrete positions.  An upper limit to the contribution
from streaming motions can be obtained measuring the difference
between the central velocity of the tidal knot and the systemic
velocity of the parent galaxy. Generally, this velocity offset is much
smaller than the small scale velocity gradient found in the tidal
objects.

\subsubsection{Outflows associated with super-winds}
Outflows are known to accelerate gas to high velocities.  Superwinds
driven outflows seem to prevail in local starbursts \citep{HAM90}, in
distant Lyman Break Galaxies \citep{PSS+01}, and in dwarf galaxies
\citep{Mar98,BWT+01}.  The kinematic signature of such winds are broad
or double-peaked emission lines.  They are due to the shell-like
structures generated by stellar winds, which always show velocity
components towards and away from the observer. If part of the shell
was shielded from view by dust, it could also produce velocity
gradients.  We did not find any obvious broadening of the emission
lines in regions characterized by a large velocity gradient and only a
moderate extinction was measured there, which makes the dust
obscuration of one component quite unlikely.

\subsubsection{Gravitationally dominated motion}
The last possible origin of the velocity gradients is gravitationally
dominated motion within an entity that became kinematically
independent from its host tail. Some of the \dV\ values measured are
certainly too large to be caused by true Keplerian rotation of a
virialized entity. If the knots were virially relaxed then one would
derive a mass of $10^{10}\,M_\odot$ for an observed gradient of
\dV$=200$\kms over 1\,kpc from the virial theorem. In those cases it
is possible that part of the velocity is due to motion of the \HII
regions towards the center of an entity in formation. The mass derived
using the Virial Theorem would then strongly overestimate the real
mass of the knot.

As the blue luminosity typically rises with mass, the velocity
gradients are expected to increase with absolute magnitude, if they
were all caused by the rotation of massive objects.  This is not
obvious in Fig.~\ref{fig:dV_MB}: we see high velocity gradients for
both faint and bright knots. However, as most of the knots are still inside
the tidal tail, show internal structure, and are not relaxed objects,
we cannot assume Keplerian rotation. Also, all objects where we could
measure velocity gradients are objects with strong emission lines,
probably sites of strong star formation, where $M_B$ or in general
optical luminosities are not a good measure for the mass of the knots.
We will further pursue this question using near infrared luminosities
in a future paper (Weilbacher et al., in prep.).  We also see from
Fig.~\ref{fig:dV_R} that the number of knots with visible gradients is
higher in the outer regions (20 to 35\,kpc) than in the inner regions
($<$20\,kpc) of our interacting systems. A reason for this could be
that decoupled kinematics or bound knots are more difficult to produce
in the denser parts of the inner tidal tails.

A good example of a self-gravitating object seems to be AM\,1159-530a:
the velocity gradient extends over 5\farcs5 or 1.6\,kpc with two
bright \HII regions at both ends. The gradient is confirmed by the
detection of absorption lines {\it between} the \HII regions with {\it
  intermediate} velocity.  Another fainter \HII region further
westward has a discrepant velocity and seems to be projected onto this
entity or falling into it.

\subsubsection{Limitations of these observations}
Clearly the large velocity gradients observed with EFOSC2 in a number
of interacting systems should be confirmed with observations at higher
spectral resolution. To unquestionably prove the amplitude of the
gradients, velocity {\it resolutions} better than 200\kms will be
required.  The system AM\,1353-272 in our sample has actually been
re-observed with FORS2 on the VLT at a higher spectral resolution
confirming velocity gradients in several knots along both tidal tails
of the interacting system, with velocity amplitudes in the range
$24<$\dV$<343$\kms \citep{WFDF02}.
To overcome the poor sampling of the gradients, imaging with higher
spatial resolution is also needed. \Ha observations with the
Fabry-Perot technique \citep[see e.g.][]{MdO+01} in some interacting
systems are in progress (Duc et al.~2002, in prep.).
%
%
\subsection{Selecting Tidal Dwarf Galaxies}
Are any of the knots present in the interacting systems studied here
the progenitors of TDGs?  We describe below the steps required to
identify a genuine TDG according to the definition cited in
Sect.~\ref{sec:introEFOSC} and check whether the objects in our sample
fulfill them. A necessary first step is to select the brightest knots
in the tidal tails with luminosities of dwarf galaxies
($-10.5\lesssim{}M_B\lesssim-17.0$\,mag, from Paper~I). Together with
the new objects presented in Table~5
we start with {\bf 44 TDG candidates} (including AM\,0529-565D).

\subsubsection{Redshift criterion}
The selection method from Paper~I to disentangle star-forming knots
along tidal features from background sources using images in three
optical broad band filters in combination with evolutionary synthesis
modeling is an efficient method (Sect.~\ref{sec:veloMeasEFOSC}).
Nevertheless, it is important to check the association with the main
interacting system by measuring the spectroscopic redshift of the
candidates. This leaves {\bf 29 candidates}.

\subsubsection{Metallicity criterion}
Low-mass companion galaxies, existing prior to the collision, can be
discriminated from TDG candidates from their low metallicities.  While
the tidal knots in our sample have almost constant metallicities
around \logOH$\approx 8.3$, three objects close to the tidal features
have abundances at least 0.4\,dex lower than the tidal knot with the
lowest abundance.  This is
puts them close to the luminosity-metallicity relationship of normal
dwarf galaxies.  A tidal origin can therefore be excluded for these
three exceptions. {\bf 26 candidates} for TDGs are left.

\subsubsection{Dynamical criterion}
Genuine TDGs should be gravitationally bound. We do not find classical
rotation curves or typical velocity dispersion in any of our TDG
candidates. Instead, we find several cases of apparent {\it velocity
  gradients} in the knots in the tidal tails. While we cannot yet
exclude all other possible causes {\bf 13 TDG candidates} seem to have
kinematics decoupled from the hosting tail. Further data with high
resolution will be necessary to prove these gradients as internal
kinematics of the individual candidates.

\subsubsection{Final sample of TDG candidates}\label{sec:resEFOSC:numlum}
The 13 objects identified as good TDG candidates represent 46\% of the
knots that were observed spectroscopically and half of the knots left
after selecting the knots depending on their metallicity.  Note that
-- given the low resolution of our spectroscopic data -- they cannot
yet be considered as real TDGs.

From our 14 interacting galaxies 5 (i.e.~about 1/3 of the sample
systems) have TDG candidates with kinematical signatures. Of those
systems, one (AM\,1054-325) has only one TDG candidate, which has an
optical appearance very similar to other knots in the same system for
which we did not detect decoupled kinematics.  The other system with
only one TDG, AM\,1159-530, has one tail with relatively high and one
with very low surface brightness. The massive TDG is positioned at the
end of the bright tail. In both galaxies with two TDGs (AM\,0537-292
and AM\,0547-272), they are positioned in the tails on opposite sides
of the nucleus. The system AM\,1353-272 contains more that half of the
TDG candidates we have finally selected, distributed over both tails.
%
%
%
\section{Conclusions}\label{sec:conclEFOSC}
We have analyzed a sample of 14 southern interacting galaxies using
deep, low resolution spectrophotometry. The knots in the tidal
features were previously identified as good TDG candidates by
comparing optical broad band imaging with evolutionary synthesis
models. The spectra were first used to investigate the association
with the interacting system by comparing the redshifts.  All knots
that were {\it detected} are actually associated with the parent
galaxies. This shows that our photometric method is a powerful tool
for preselecting TDGs. The few knots which were not detected in our
spectra could be of a different nature.

We determined the oxygen abundances of the knots and found a mean
value of \logOH$=8.34$ with a scatter of $\pm$0.2\,dex. Both the mean
value and the scatter are in good agreement with those found in the
outer disks of spiral galaxies, which precludes the possibility that
these knots could be pre-existing objects infalling onto the
interacting systems from outside. Three dwarf galaxies close to the
tidal features are unlikely to be TDGs due to their low metallicity.

As predicted by our photometric models, the knots are actively forming
stars, and some knots may even be starbursts with high \Ha
luminosities.  \LHa of the knots within our spectroscopic slits is
higher by two orders of magnitude than that of typical individual \HII
regions in spiral disks and lower than that of local star-forming
galaxies by a factor of 10. However, the flux represents many
individual \HII regions whose number depends on the slit width and the
distance of the interacting system.
Due to the overlap of the distribution of \LHa with ``normal'' giant
\HII regions in spiral disks and the difference to dwarf galaxy
luminosities, \LHa is not a good criterion to discriminate TDGs from
other phenomena.

We see surprisingly strong velocity differences in some interacting
galaxies, and 13 TDG candidates show kinematic signatures which could
be internal velocity gradients in these knots. The velocity amplitudes
are in some cases more than one order of magnitude higher than those
observed in spiral disks. Our analysis and interpretation is hampered
by the small angular size of these gradients, which are barely
resolved in $\sim$1\arcsec\ seeing, so we could not discriminate
possible decoupled kinematics from other phenomena like chance
alignment of multiple bodies which mimic velocity gradients. We can
only prove the existence of one strong velocity gradient within the
TDG AM\,1159-530a, where we see the gradient extending over 5\farcs5,
which is likely due to the gravitational forces within the body of the
TDG.  To confirm the velocity gradients of the knots in the tidal
tails as decoupled kinematics governed by internal gravitation an
improvement of at least a factor two in spatial resolution is
required. Higher velocity resolution (better than 200\kms) would be
desirable to confirm the amplitude of the velocity gradients.  The
discussion of both emission line luminosities and velocity gradients
is restricted by the nature of our multi-slit spectroscopic data,
which allows us to analyze only discrete parts of the whole
interacting system. Narrow band imaging would be required to confirm
and interpret the full range of \Ha luminosities found in the TDG
candidates. Furthermore, observational methods like Fabry-Perot- or
integral field spectroscopy would be required to fully analyze the
velocity distributions of the ionized gas in the interacting galaxies
and the TDG candidates.

Finally, the data collected in this paper will help us refine our
spectrophotometric evolutionary synthesis models for analyzing the
starburst properties of the TDG candidates and to derive more accurate
predictions of their future evolution.

\begin{acknowledgements}
  We would like to thank L.-M.~Cair\'os for careful reading and helpful 
  comments on the manuscript, and R.~Hessman who was very helpful in
  improving the English.
  Many thanks to Pierre Leisy for exceptionally dedicated and helpful
  support during the observing run in January 2000 on La Silla.
  PMW acknowledges partial support from DFG grants FR 916/6-1 and FR
  916/6-2.
\end{acknowledgements}
\bibliography{/home/leo/weilbach/Texte/PmW}
\cleardoublepage
%
%
\end{document}